\documentclass[traditabstract]{aa}   
\usepackage{graphicx}
\usepackage{txfonts}
\usepackage{color}
\usepackage{natbib}
\usepackage{epsfig}
\usepackage{hyperref}
\usepackage{longtable}
\usepackage[switch]{lineno}

\begin{document}
   \title{Connection between inner jet kinematics and broadband flux variability in the BL Lac object S5 0716+714}

   \subtitle{ }

   \author{
          B. Rani \inst{1}
          \and T. P.\ Krichbaum \inst{1}
          \and A. P. Marscher \inst{2}
          \and J. A. Hodgson \inst{1}
          \and L. Fuhrmann \inst{1}
          \and E. Angelakis  \inst{1}
          \and S. Britzen  \inst{1}
          \and J. A. Zensus \inst{1} 
          }

   \institute{ 
              Max-Planck-Institut f{\"u}r Radioastronomie (MPIfR), Auf dem H{\"u}gel 69, D-53121 Bonn, Germany 
          \and 
            Institute for Astrophysical Research, Boston University, 725 Commonwealth Avenue, Boston, MA 02215, USA
           }

   \date{Received ---------; accepted ----------}

  \abstract
{ We present a high-frequency very long baseline interferometry (VLBI) kinematical study of the 
BL Lac object S5 0716+714 over the time period of September 2008 to October 2010. The aim of the study 
is to investigate  the relation of the jet kinematics to the observed broadband flux variability. We find significant 
non-radial motions in the jet outflow of the source. In the radial direction, the highest measured apparent speed is 
$\sim$37~c, which is exceptionally high, especially for a BL Lac object. Patterns in the jet flow reveal a roughly 
stationary feature $\sim$0.15~mas downstream of the core. The long-term fits to the component 
trajectories reveal acceleration in the sub-mas region of the jet. The measured brightness temperature, T$_{B}$, 
follows a continuous trend of decline with distance, T$_B \propto r_{jet}^{-(2.36\pm0.41)}$, which suggests a gradient 
in Doppler factor along the jet axis. 
Our analysis suggest that a moving disturbance (or a shock wave) from the base of the jet produces the 
high-energy (optical to $\gamma$-ray) variations upstream of the 7~mm core, and then later causes 
an outburst in the core. Repetitive optical/$\gamma$-ray flares and the curved trajectories of the 
associated components suggest that the shock front propagates along a bent trajectory or helical path. 
Sharper $\gamma$-ray flares could be related to the passage of moving disturbances 
through the stationary feature.  Our analysis suggests that the $\gamma$-ray and radio emission regions have 
different Doppler factors. 
}



\titlerunning{Jet kinematics and high-energy emission connection in 0716+714}
\maketitle

\section{Introduction}
The combination of very long baseline interferometry (VLBI) images and broadband flux density variability is a unique way to 
probe the emission mechanisms near the base of jets in the blazar class of active galactic nuclei (AGN). 
The broadband flares are often found to be connected to the ejection of new moving emission components into
the jet \citep[e.g.,][and references therein]{marscher2008, jorstad2013, schinzel2012, krichbaum2001}. 
Moreover, mm-VLBI observations offer a unique possibility to study the structural evolution in 
the parsec-scale region, which has been proposed to be the site of much of the high-energy emission 
\citep[e.g.,][]{rani2013_3c273, rani2014, lars2014, marscher2008, schinzel2012}. Therefore, these observations have provided new 
constraints on the physical parameters of the emission regions, i.e., size, brightness temperature, magnetic field, and motion.  

The blazar S5 0716+714 \citep[z $\sim$ 0.3,][]{nilsson2008, danforth2012} is a BL Lac object with a featureless 
optical spectrum. It is one of the most intensively studied blazars because of its extreme variability 
properties across the entire electromagnetic spectrum \citep[e.g.,][]{villata2008, fuhrmann2008, rani2010a, rani2010b, rani2013a, 
rani2013b, larinov2013}.  The broadband flux variability of the source is quite complex, with rapid flaring activity (on a timescale of a few hours to days) superimposed on top of a 
broad and slow variability trend on a timescale of $\sim 1$ year \citep{rani2013a, raiteri2003}. VLBI studies of the source show a 
core-dominated jet pointing towards the north 
\citep{bach2005, britzen2009}, while Very Large Array observations show a halo-like jet misaligned by $\sim$90$^{\circ}$ on kiloparsec scales. 
\citet{britzen2009} suggested an apparent stationarity of jet components relative to the core;
more recent 
studies, however, have reported motion as fast as $\sim$40~c \citep{rastorgueva2011, larinov2013, lister2013}. Non-radial motion and 
wiggling component trajectories have often been observed in the inner mas jet region of the 
source \citep{britzen2009, rastorgueva2011, rani2014}. 

In \citet{rani2013a} (hereafter Paper I), we presented the densely sampled multi-frequency observations of the source 
between April 2007 and January 2011. These observations allowed us to study the broadband flaring behavior 
of the source and to probe the physical processes, location, and size of the emission regions. The intense optical and $\gamma$-ray 
monitoring revealed fast repetitive variations (60--70 days) superimposed on a long-term variability 
trend with a time scale of $\sim$350 days, which propagated down to radio wavelengths with an observed time lag of $\sim$65 days. 
A detailed investigation of the optical flares found a variability amplitude proportional to the flux level, which 
can be explained by a variable Doppler factor. We also found that the shock-in-jet model for the evolution of radio flares
requires geometrical variations in addition to intrinsic variations of the source. 
As a possible scenario to explain the observations, we suggested that the geometry significantly affects the 
long-term flux variations, which could be caused by a relativistic shock tracing a spiral path through 
the jet. To explain the multi-frequency behavior of an optical-$\gamma$-ray outburst in 2011, \citet{larinov2013} have also suggested 
a shock wave propagating along a helical path in the blazar's jet. 

In this paper, we use high-resolution multi-frequency VLBI observations to investigate the inner jet kinematics of S5 0716+714,
with a focus on the major radio/optical/$\gamma$-ray flares over the time 
period of September 2008 to October 2010.  We focus on the morphological evolution of 
the source to investigate its relation to the broadband flux variations reported in Paper I, in particular the high-energy 
emission. The paper is structured as follows. Section 2 provides a brief
description of observations and data reduction. In Section 3, we
report and discuss our results. Summary and conclusions are given in Section 4.



\section{Multi-frequency VLBI data: observations and data reduction}
To explore the inner jet kinematics of the source, we used the 7~mm (43~GHz) and 3~mm (86~GHz) VLBI data 
obtained between September 2008 and October 2010, with observations at 26 epochs during this period.
The 7~mm data are from the 
Boston University monthly monitoring program of bright $\gamma$-ray blazars with the Very Long Baseline Array (VLBA)\footnote{VLBA-BU-BLAZARS, http://www.bu.edu/blazars}. 
The data reduction was performed using 
standard tasks of the Astronomical Image Processing System (AIPS) and Difmap \citep{shepherd1997} software.
The imaging of the source (including amplitude and phase self-calibration) was done using the {\it CLEAN} 
algorithm \citep{hogbom1974} and {\it SELFCAL} procedures in Difmap \citep{shepherd1997}. 
Several iterations of phase corrections followed by amplitude adjustments were adopted for the self-calibration 
process. More details of the data reduction can be found in \citet{jorstad2005}.

The 3~mm observations were performed using the Global mm VLBI Array (GMVA). The GMVA is currently an array consisting 
of 14 antennas in Europe and the United States, including the 8 VLBA stations equipped with 3~mm receivers, plus
the Effelsberg, Onsala, Mets{\"a}hovi, Pico Valeta, Plateau de Bure, and Yebes antennas. 
Between October 2008 and May 2010, observations were taken approximately every six months. Amplitude calibration 
is difficult at 3~mm, with atmospheric fluctuations often having a major effect. We find that amplitudes are frequently too low 
after the nominal calibration in AIPS. To correct for this, amplitudes are scaled by comparing the flux density on the shortest baselines with 
single dish measurements. Details of the 3~mm data reduction can be found in Hodgson et al. (2014).

\begin{figure*}[t]
   \centering
\includegraphics[scale=0.7,angle=0]{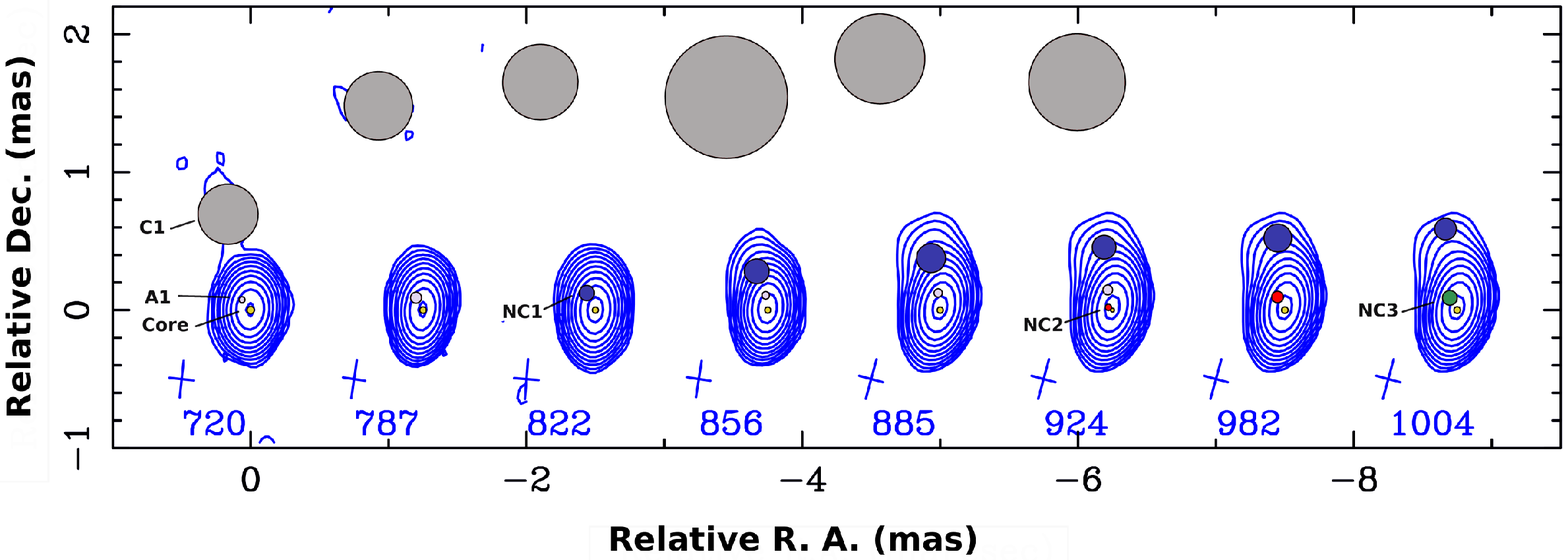}
 \includegraphics[scale=0.7,angle=0]{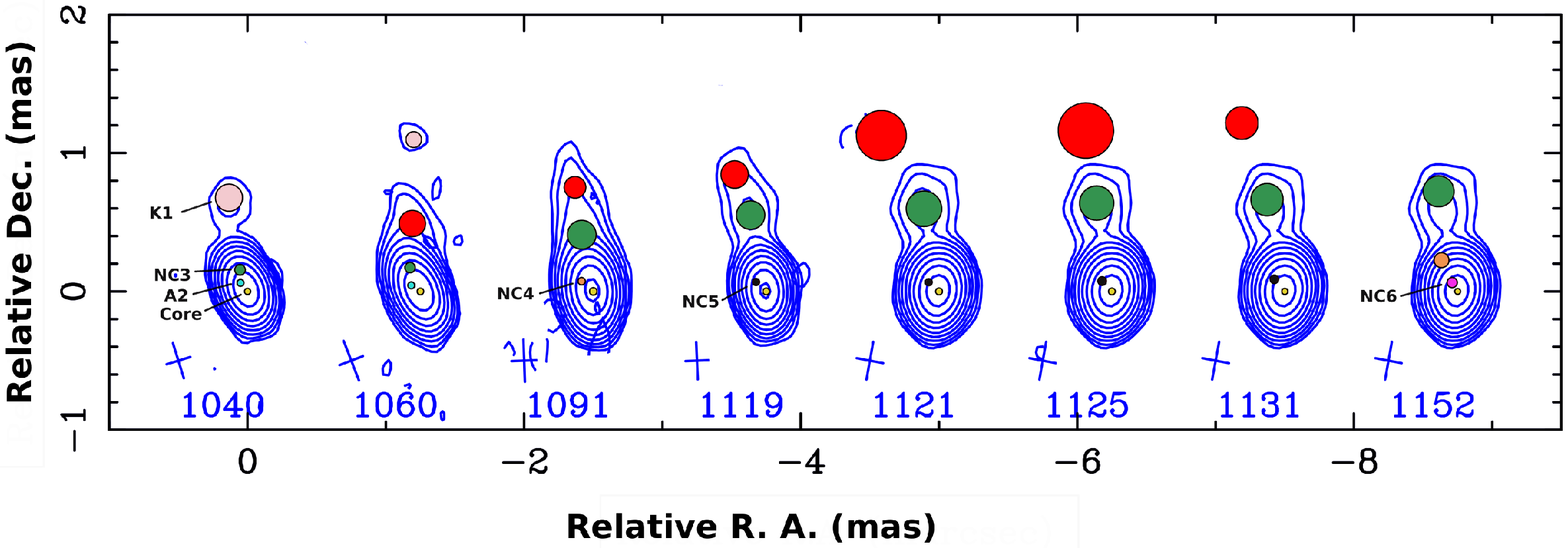}
 \includegraphics[scale=0.7,angle=0]{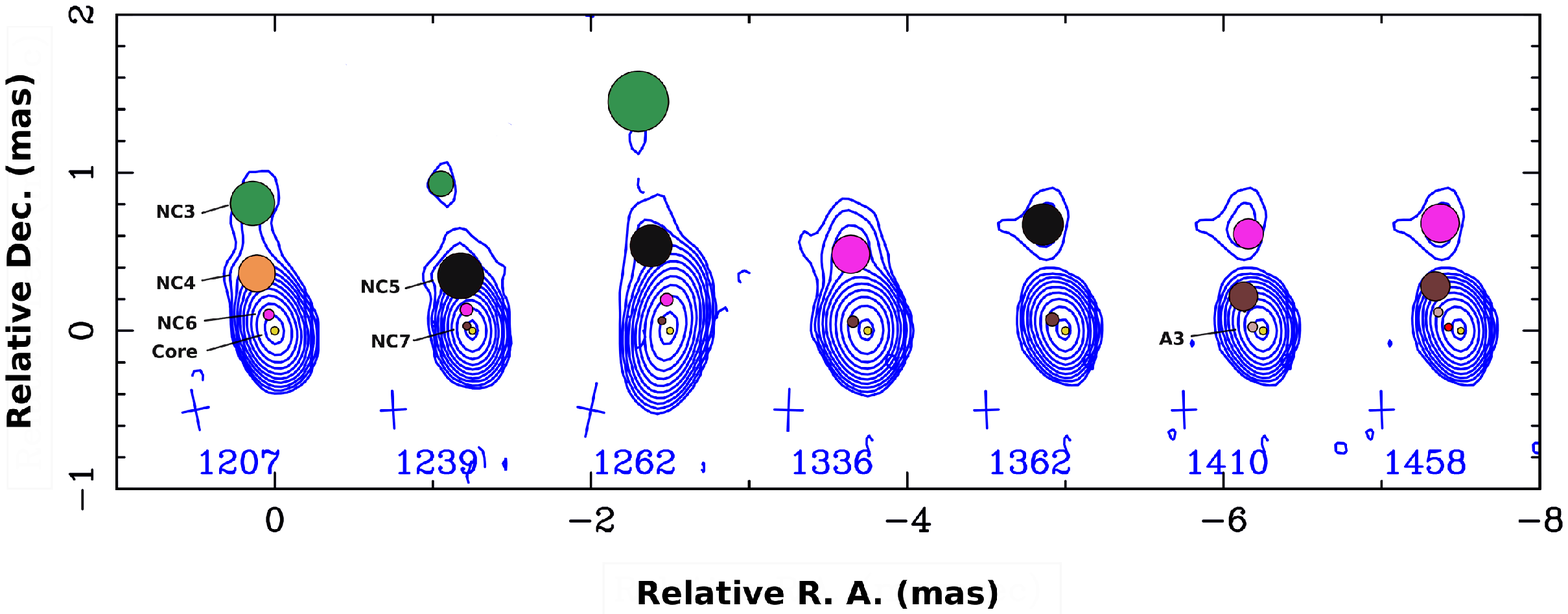}
  \caption{43 GHz VLBA images (contours) and Gaussian components of S5 0716+714 at different epochs between September 
2008 and October 2010. The date of observation is given at the bottom of each map (shifted Julian day, JD$-$2454000). 
The fitted model parameters for each epoch are given in Table \ref{tab1}.  
The cross at the bottom of each map represents the FWHM of the restoring beam (0.25$\times$0.18 mas). 
Contour levels correspond to -0.003 0.003 0.006 0.012 0.024 0.048 0.096 0.19 0.38 0.76 1.54 3.07 6.14 12.28 24.58 
49.15 98.30 mJy/beam.               }
\label{plot_fig1}
\end{figure*}

\begin{figure*}
   \centering
\includegraphics[scale=0.7,angle=0]{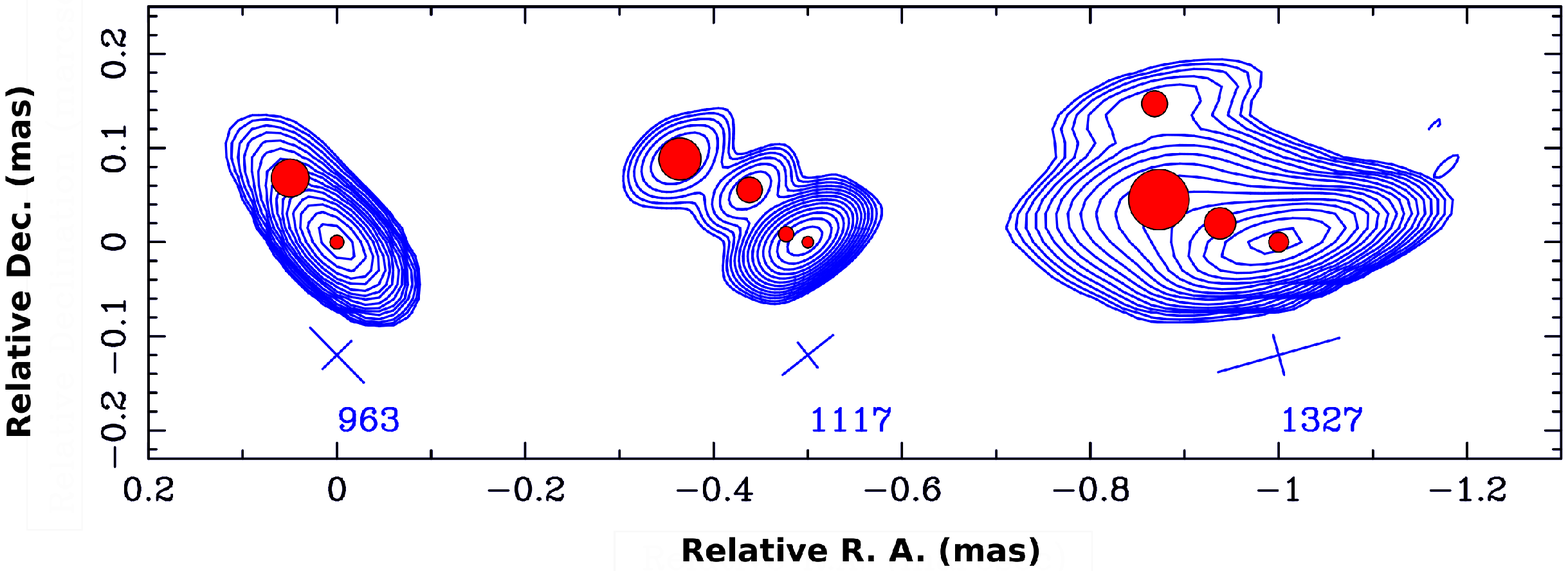}
  \caption{The inner jet structure of S5 0716+714 from GMVA observations at 3 mm. The cross at the bottom of each 
map represents the FWHM of the restoring beam (0.07$\times$0.04 mas). Contour levels 
correspond to -0.004 0.004 0.008 0.016 0.032 0.064 0.12 0.25 0.51 1.02 2.04 4.09 8.19 16.38 
32.77 65.54 mJy/beam.           }
\label{plot_fig2}
\end{figure*}

We modeled the observed brightness distribution (the visibility amplitude and phases of the observed 
radio brightness of the source) by multiple circular Gaussian components, thus deriving positions, flux 
densities, and sizes of the distinct bright features in the jet. The model fits were carried out for all 
epochs within the DIFMAP package, starting with a point-like model and fixing the position of the brightest 
component to (0,0). Jet components were added until the addition of an extra component did not lead to a 
significant improvement in the $\chi^2$ value of the fit to the uv data. The uncertainties of the model 
component parameters were determined by comparing the parameter ranges obtained after performing model fits 
with a different number of model components. We used at least four different model fits for each epoch to 
obtain the parameter uncertainties. However, the uncertainties depend also on the self-calibration and data 
editing, and on the brightness and size of the components. 
In addition, the uncertainties increase with increasing distance from the core.  This is accounted 
for by following
an independent approach for the error estimation proposed by \citet{krichbaum1998}. The fitted model 
parameters for all of the
epochs are listed in Table \ref{tab1}, where we only list the formal errors obtained 
from the first method. Figures \ref{plot_fig1} and \ref{plot_fig2} show the fitted circular Gaussian components superposed on the clean images.

To investigate the kinematics in the jet of S5 0716+714, the individual model components were identified  
following the assumption that the changes of the flux density, distance from the VLBI core, position angle, and 
size should be small for the time period between adjacent epochs. 
In order to prevent a potentially large systematic error arising from the incorrect cross-identification of moving 
features from epoch to epoch, the simplest scheme was adopted while identifying the jet-features. A 
self-consistent cross-identification is proposed using all available model-fit parameters.  
These cross-identifications are not necessarily unique, especially if the source evolves in a more complex manner than assumed here.


\section{Results and discussion} 
In this section we present the flux and spatial evolution of the bright radio emission in the jet 
of S5 0716+714. Special attention is given to the kinematics of the inner jet ($<$\,2~mas from the core).  
We also investigate possible correlations between the jet kinematics
and the broadband flux variations reported in Paper I.

\subsection{Jet kinematics} 
Figures \ref{plot_fig1} and \ref{plot_fig2} display a sequence of VLBI maps convolved with a natural beam. 
The model fit parameters are given in Table \ref{tab1}. In the following sub-sections, we will discuss 
in detail the apparent motions of components and brightness temperature gradient in the parsec-scale jet.

\subsubsection{Component motion}
\label{comp_motion}
\begin{figure}
   \centering
\includegraphics[scale=0.33,angle=0, trim=0 0 0 0.5, clip]{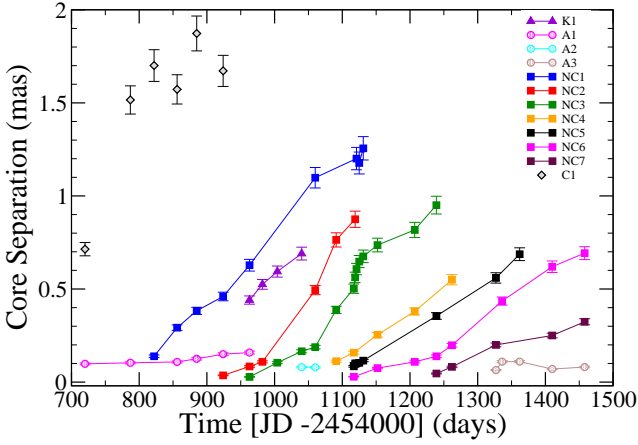}
  \caption{Evolution of core separation of the individual components as a function of time.
              }
\label{plot_fig5}
\end{figure}

\begin{figure}
   \centering
\includegraphics[scale=0.48,angle=0, trim=0 0 0 0, clip]{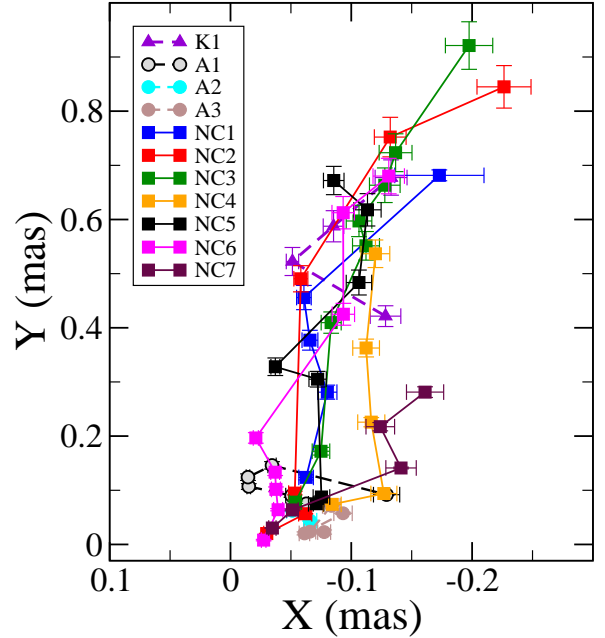}
  \caption{Trajectories of the jet knots in the X-Y plane,
where X = r.cos($\mathrm{\theta}$ + 90) and Y = r.sin($\mathrm{\theta}$ + 90) (r is the radial separation from the core
and ${\rm \theta}$ is the position angle with respect to an imaginary north-south line drawn through the map center).
              }
\label{plot_fig6}
\end{figure}

For the kinematic study of individual components, we choose the VLBI core as a reference point, 
fixed to coordinates (0,0). However, it is important to note that the absolute position of the 
reference point i.e.\ core may change e.g.\ due to changes in opacity or instabilities in  the jet; but, 
the results presented here are not sensitive to that as we are interested in the relative motion of components 
w.r.t.\ the core. The VLBI data were fitted by using different numbers of components 
during different activity stages over the time span considered here. During 
this period, we identified a total of 12 components - C1, K1, A1, A2, A3, and NC1 to NC7 - in addition to 
the core, C0 (see Fig. \ref{plot_fig1}). 
Figure \ref{plot_fig5} plots the evolution of the distance of different knots from the core, and 
Fig. \ref{plot_fig6} shows their trajectories in the XY-plane projected on the sky.  \\

\noindent 
{\bf Stationary Features :} \\
The sequence of VLBI images allow us to investigate the component motion along the jet as a function 
of time. Most of the components exhibit significant motion down the jet, except A1, A2, and A3. 
The latter are comparatively stable in position, although they exhibit significant scatter 
in position angle exceeding the uncertainties (see Fig.\ \ref{plot_fig6}). Nevertheless, their radial 
distance from the core remains at $\sim$0.15 mas. Stationary features are 
a common characteristic in AGN jets \citep[e.g.,][]{fromm2013, jorstad2001, britzen2010}. In straight jets, 
these can be produced by recollimation shocks, instabilities [magnetohydrodynamic (MHD), and/or Kelvin-Helmholtz (KH)], or 
magnetic pinches \citep[see][for details]{marscher2009, hardee2006, meier2001}. Bends in the jet can also 
cause quasi-stationary features, either because the jet turns more into the line of sight, thus increasing the Doppler beaming factor, or due to the 
formation of a shock that deflects the flow \citep{alberdi1993}. Numerical simulations indicate that when a moving knot passes through 
a standing re-collimation shock, the components blend into a single feature, then split up after the collision 
with no lasting changes in the proper motion of the moving knot \citep{gomez1997, fromm2012}. During the interaction, the feature may move a short distance downstream before returning to its previous position. Therefore, it is quite possible 
that A1, A2, and A3 are the same standing feature, which becomes disturbed by moving features. We find that the kinematics 
of three new components NC2, NC3, and NC6, are significantly different before and after the interaction. The 
components move more slowly ($\beta_{app} \leq 10c$) as they approach the stationary feature, then 
faster ($\beta_{app} \geq 20c$) after the interaction. Such behavior is consistent with the stationary feature representing a bend 
in the jet \citep{alberdi1993}. Indeed, a bend at $\sim$0.15~mas is evident in the 3~mm maps of the source  
(see Fig.\ \ref{plot_fig2}). This suggests that the observed stationary features are standing oblique shocks that
cause the flow of the jet to bend at $\sim$0.15~mas owing to a transverse pressure gradient or an impact with an interstellar cloud. \\

\noindent 
{\bf Moving Features :} \\
The other components, C1, K1, and NC1 to NC7, exhibited significant motion in the radial direction. Component 
C1 represents very faint emission at a core separation of $\geq$1 mas. It seems to be the remnants of a component 
ejected in earlier epochs. The bright feature K1 could be a trailing component that forms behind a strong shock 
\citep{agudo2001} as it is identified at four epochs after the ejection of a new component (NC1).

As seen in Fig. \ref{plot_fig6}, all components follow curved trajectories in the XY-plane projected on the sky. 
In addition, the trajectories differ from one component to the next. 
The wiggling trajectories in the jet might be a signature of helical motion. 
Helical jet models \citep[e.g.,][]{gomez1994, hardee2006} represent the jet as an inhomogeneous flow of
plasma interacting with ambient matter. 
Magnetohydrodynamic or Kelvin-Helmholtz instabilities \citep{hardee2006, prucho2006} 
can explain the bending and helical structures, and possibly also less regular
wiggles observed in parsec-scale radio jets.

\begin{figure}
   \centering
\includegraphics[scale=0.35,angle=0, trim=0 0 0 0.5, clip]{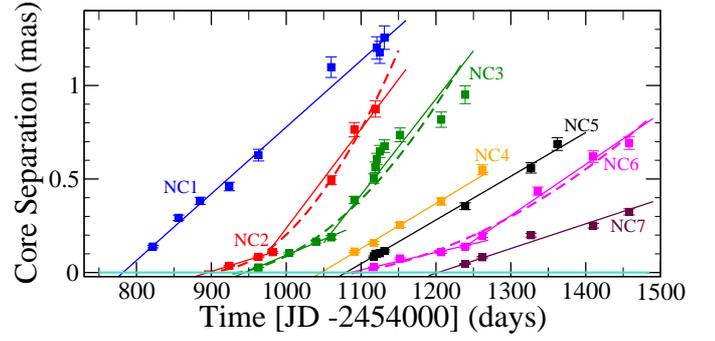}
  \caption{Evolution of radial separation of different components from the core. The solid lines 
represent fits with linear functions, while the dashed curves are quadratic fits.               }
\label{plot_fig7}
\end{figure}

In order to investigate component motions in the jet of S5 0716+714, we use the XY positions of the 
components relative to the core to fit the trajectories with polynomials of different order. The 
resulting fitted radial separation of each component, $r(t) = \sqrt{x(t)^2 + y(t)^2}$, is shown in Fig. \ref{plot_fig7}. We 
find that a linear function is sufficient 
to  fit the trajectories of components NC1, NC4, NC5, and NC7. However, a second-order polynomial is required to 
fit the trajectories of components NC2, NC3, and NC6, which corresponds to an apparent acceleration. Two multiple straight lines needed to fit different portions of the trajectories of these components in Fig. \ref{plot_fig7} provides a
clear demonstration of the inadequacy of simple linear fits.  The dashed curves represent the quadratic fits.

For each component, the fits yield an average proper motion and a mean speed. 
Using the estimated angular speed $\mu$, we have computed the kinematic parameters of the jet, e.g., the apparent
speed, $\beta_{app}$, and Doppler factor, $\delta$. We derive the apparent speed of the
components, $\beta_{app}$, from the angular speed, $\mu$, using
\begin{equation}
\beta_{app} = \frac{\mu~ d_L}{c(1+z)},
\end{equation}
where $d_L$ is the luminosity distance and $z$ the redshift of the source. The luminosity distance $d_L$ corresponding 
to z=0.3 is $d_L$ = 1600 Mpc for a $\Lambda$CDM cosmology with $\Omega_m$ = 0.27, $\Omega_{\lambda}$ = 0.73, and
$H_0$ = 71 km s$^{-1}$ Mpc$^{-1}$ \citep{spergel2003}. The angular and apparent speed of the 
individual components are given in Table \ref{tab2}. The calculated $\beta_{app}$ values range from $6c$ to $37c$, the latter of which is unusually high for a blazar, especially for a BL~Lac object.

For the non-ballistic components, NC2, NC3, and NC6, the quadratic fits represent significant acceleration along the jet axis.
Formally, we obtain $\dot{\mu}$ = 4.90$\pm$1.55 mas yr$^{-2}$ for NC2, 
$\dot{\mu}$ = 2.45$\pm$0.80 mas yr$^{-2}$ for NC3, and $\dot{\mu}$ = 0.94$\pm$0.36 mas yr$^{-2}$ for NC6.
Such apparent acceleration is a commonly 
observed feature of many objects in the MOJAVE sample \citep{lister2013}, for which a statistically significant 
tendency for acceleration in the base of jets has been found.
A standing shock slows the flow while deflecting it, after which the flow can accelerate back to
the velocity it had upstream of the shock
\citep{gomez1997}. However, since the apparent speed was relatively slow in the upstream region, this seems not to fit the case of S5 0716+714.
The data are more consistent with the scenario that, after a moving component passes through the stationary feature, it
bends towards or away from our line-of-sight. It then appears to move faster because its velocity vector subtends an angle closer to that which maximizes apparent motion, $\arccos(\beta)$, where
$\beta$ is the velocity of the component divided by the speed of light.
An alternative possibility is that the underlying flow of the jet systematically accelerates outward.
Theoretical models involve strong magnetic fields associated
with the putative supermassive black hole/accretion disk system that play a key role in the initial acceleration 
and collimation of the jet \citep{blandford1982, blandford1977, meier2001}. The conversion 
of Poynting flux to flow energy is gradual and may persist out to parsec scales \citep[e.g.,][]{sikora2005}.
Under this explanation, however, it is difficult to explain the straight-line kinematic evolution of some of the components. \\

\noindent 
{\bf Variability Doppler factor:} \\
A variability Doppler factor using VLBI jet kinematics was defined by \citet{jorstad2005} as  
\begin{equation}
\delta_{VLBI} = \frac{s d_L}{c \Delta t_{var} (1+z)},
\end{equation}
where $s$ is the angular size of the component, defined as 1.6$\times$$\phi$ for a Gaussian with FWHM diameter $\phi$ measured at 
the epoch of maximum brightness, and $d_L$ is the luminosity distance. The timescale of variability is defined as 
$\Delta t_{var}$ = $dt/ln(S_{max}/S_{min})$ \citep{burbidge1974}, where $dt$ is the time separation in years between 
maximum ($S_{max}$) and minimum ($S_{min}$) flux densities.  The estimated values of $\delta_{VLBI}$ for the 
individual components, listed in Table \ref{tab2}, range from 6 to 21. 
Using the broadband flux density and spectral variability study of the source over the same time period (Paper I, Sect.\ 3.4), 
we obtain a range of self-consistent values for the Doppler factor. The different independent approaches suggest that 
$\delta \geq$20, which is consistent with upper range of the values of $\delta_{VLBI}$ obtained here.

\subsubsection{Ejection Epochs}
Back-extrapolation of the components' motion allows us to estimate the time of zero separation 
from the core (i.e., the ``ejection'' time, T$_0$). As shown in the previous section, the 
components follow curved trajectories; therefore, we back-extrapolate the component trajectories in 
the XY-plane to determine the epoch of (0,0) separation. The error estimates are not very straightforward 
to calculate for components following non-ballistic trajectories. We use the estimated error on the 
fitted parameters for a given function to calculate its uncertainty. Back-extrapolation of the 
two envelopes of the function provide us with the error in T$_0$. The calculated ejection times 
for the individual components are listed in Table \ref{tab2}.

\begin{table}
\center
\caption{Physical parameters of the radio emission}
\begin{tabular}{l c c c c } \hline
Component  & $\mu$ (mas/yr)  & $\beta_{app} (c)$ & $\delta_{var}$  &T$_0$ (days)        \\\hline \smallskip
NC1        &1.30$\pm$0.04  &  24.88$\pm$0.88  & 7.03$\pm$0.35      &783$^{+16}_{-12}$   \\\smallskip
NC2$_l$    &0.45$\pm$0.00  &   8.71$\pm$0.13  & 13.84$\pm$0.69     &900$^{+10}_{-16}$  \\\smallskip
NC2$_u$    &1.93$\pm$0.14  &  37.06$\pm$2.75  &                    &                   \\\smallskip
NC3$_l$    &0.61$\pm$0.03  &  11.69$\pm$0.64  &  6.02$\pm$0.30     &933$^{+19}_{-12}$  \\\smallskip
NC3$_u$    &1.87$\pm$0.17  &  35.78$\pm$3.29  &                    &                   \\\smallskip
NC4        &0.87$\pm$0.04  &  16.81$\pm$0.95  &  6.91$\pm$0.34     &1047$^{+12}_{-9}$  \\\smallskip
NC5        &0.85$\pm$0.03  &  16.30$\pm$0.59  & 11.72$\pm$0.58     &1079$^{+5}_{-6}$   \\\smallskip
NC6$_l$    &0.34$\pm$0.04  &   6.53$\pm$0.77  & 15.33$\pm$0.76     &1092$^{+26}_{-17}$ \\\smallskip
NC6$_u$    &0.99$\pm$0.04  &  19.05$\pm$0.92  &                    &                   \\\smallskip
NC7        &0.48$\pm$0.04  &   9.20$\pm$0.84  & 20.96$\pm$1.04     &1203$^{+11}_{-16}$   \\\hline 
\end{tabular} \\
$l$ and $u$ denote, respectively, lower and upper values of the component speeds. 
\label{tab2}
\end{table}

\subsubsection{Brightness temperature gradient in the jet}
The redshift-corrected brightness temperature (T$_{B,obs}$) of the bright emission features can be
approximated using the following relation \citep{jorstad2005}:
\begin{equation}
T_{B,obs} = 1.22 \times 10^{12} \frac{S_{comp}}{(1+z)~d_{comp}^{2} ~\nu^{2}}\, ~ {\rm K},
\end{equation}
where $S_{comp}$ is the component flux density in Jy, $d_{comp}$ is the FWHM of the
circular Gaussian component in mas, and $\nu$ is the observing frequency in GHz; for the calculations 
here, we have only used 43~GHz data. The calculated brightness temperatures ($T_{B,obs}$) of the components 
generally decline with radial separation from the core (see Fig.\ \ref{plot_fig8}). This decaying 
trend can be approximated by a power law, $T_{B,obs} \sim r_{jet}^{-\epsilon}$ with $\epsilon$ = 2.43$\pm$0.36, where $r_{jet}$ is the 
radial distance along the jet axis. The solid line 
in Fig.\ \ref{plot_fig8} represents the fitted power law.

  \begin{figure}
   \centering
\includegraphics[scale=0.35,angle=0, trim=0 0 0 0.5, clip]{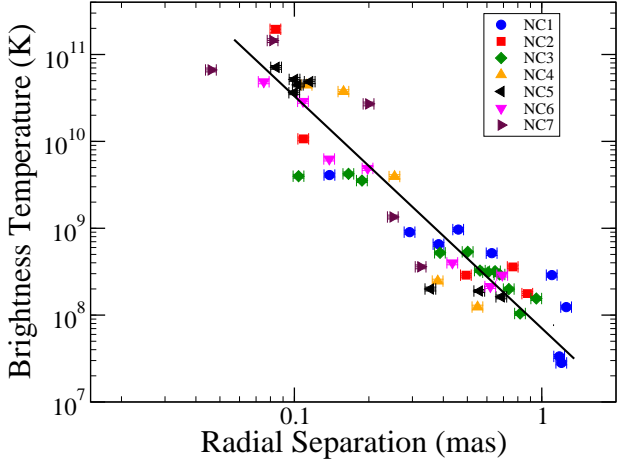}
  \caption{Component brightness temperature as a function of radial separation from the core. The solid line 
represents the fitted power law with index $\epsilon = 2.43\pm0.36$.                }
\label{plot_fig8}
\end{figure}

We place the observed brightness temperature gradient in the context of the shock-in-jet 
model \citep{marscher1985}. In the common picture of this model, a relativistic shock 
propagates down a conical jet, slowly expanding adiabatically while maintaining shock conditions during 
propagation. While the shock propagates down the jet, it undergoes three major evolutionary stages 
dominated by inverse Compton, synchrotron, and adiabatic energy losses. As a result, the observed brightness 
temperature declines as a power law, T$_{B,obs}$ $\propto r_{jet}^{-\epsilon}$. The value of $\epsilon$ can be derived from the spectral 
evolution of the radio emission \citep{lobanov1999, fromm2013_vlbi}. Following 
$T_B \propto$ $S_{\nu}r_{jet}^{-2} \nu^{-2}$ and S($\nu$) $\propto r_{jet}^{\xi} \nu^{\zeta} \delta^{(s+3)/2}$ 
\citep{lobanov1999}, we obtain  
\begin{equation}
\epsilon_{Compton} = \frac{11-s-a(s+1)}{8} + \frac{b(s+3)}{2} - 2, 
\end{equation}
\begin{equation}
\epsilon_{synchrotron} = \frac{-[4(s-1)+3a(s+1)]}{6} + \frac{b(s+3)}{2} - 2,
\end{equation}
and
\begin{equation}
\epsilon_{adiabatic} = \frac{2(5-2s)-3a(s+1)}{6} + \frac{b(s+3)}{2} - 2,
\end{equation}
where $a$, $b$, and $s$ parametrize the variations along the jet axis
of the Doppler factor, $\delta \propto r_{jet}^b$, magnetic field, $B \propto r_{jet}^{-a}$, and power law distribution of 
energy of the emitting electrons, $N(\gamma) \propto \gamma^{-s}$.
Assuming a constant Doppler factor, i.e., $b$ = 0, and using a typical value of $s$ = 2 (corresponding to optically thin 
synchrotron spectral index = $-$0.5), and $a$ = 1 to 2 (for toroidal and poloidal magnetic fields, respectively), we obtain 
$\epsilon_{Compton}$ = 1.3 to 1.6, $\epsilon_{synchrotron}$ = 4.2 to 5.7, and $\epsilon_{adiabatic}$ = 3.2 to 4.7. 

The observed brightness temperature gradient in S5 0716+714 (Fig. \ref{plot_fig8}) has a slope equal to 2.43$\pm$0.36. 
The slope of the observed intensity gradient rules out the simple assumptions of a constant Doppler factor. We therefore 
consider a variable Doppler factor along the jet axis. We find that the observed value of $\epsilon$ is consistent with 
the adiabatic loss phase if $\delta \propto$ $r_{jet}^{b}$, with $b$ $\sim$ 0.12 to 0.40, which corresponds to a moderate variation in the Doppler factor. 
However, with relatively larger variations in the Doppler factor $b$ $\sim$ 0.47 to 0.74, the 
observed value of $\epsilon$ would agree with synchrotron loss phase as well.

The evolution of the flux density and frequency ($S_m$ and $\nu_m$) at the spectral turnover from synchrotron self-absorption of the radio flares, discussed in 
Paper I (Section 3.3.2), also suggests a variation of $\delta$ along the 
jet axis. In that study, we found that $\delta$ changes as $r_{jet}^b$ with $b = 0.7$ during the rise and $b = 2.6$ during 
the decay of the first radio flare. The evolution of the second flare was governed 
by $b = 0.4$ during the rising phase and $b = -2.0$  during the decay 
(see Paper I for details). 
Therefore, it is evident that the two flares require a different dependence of $\delta$ on distance from the core, which is also inconsistent  
with what we obtained for the brightness temperature gradient ($b = 0.12$ to 0.74). 
This can be explained in two ways. First, the single-dish radio flux correlates well with the VLBI core 
flux (see Section \ref{vlbi_flare_corr} 
for details). This implies that the radio flares originate within the core. The core is more compact and farther upstream than the 
jet components, with a comparatively higher brightness temperature, and can follow a different 
dependence of $\delta$ on $r_{jet}$. The individual components trace the intensity gradient further downstream of the 
jet, and therefore can individually have a different  $\delta$ -- $r_{jet}$ dependence. A second possibility is
that the brightness temperature gradient plot corresponds to an averaged behavior of the two radio flares.

  \begin{figure*}
   \centering
\includegraphics[scale=1,angle=0, trim=0 0 0 0, clip]{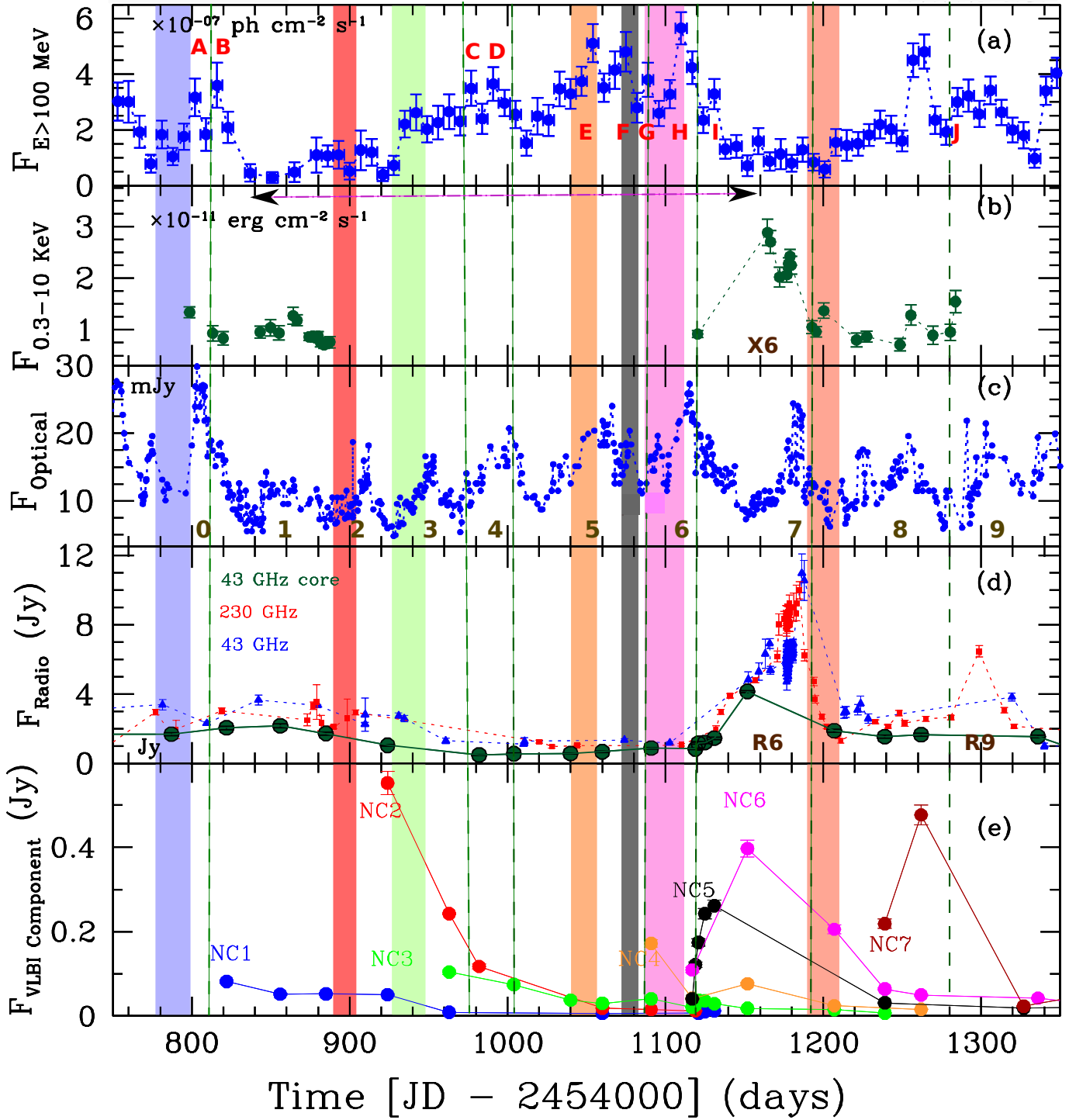}
  \caption{Broadband flux light curves of S5 0716+714: (a) weekly averaged $\gamma$-ray flux light 
curve at E$>$100~MeV, (b) X-ray light curve at 0.3--10 keV, (c) optical V-band light curve, (d) 
single-dish (43 and 230~GHz) and 7~mm VLBI core flux light curves, and (e) flux density curves of moving 
jet components shown in different colors; the same color is used to shade their respective ejection times, i.e., epochs of zero 
separation from the core. The green dashed lines indicate the average times when moving components 
cross the stationary feature at $\sim$0.15~mas.  The major outburst is marked with a horizontal arrow; the faster 
repetitive optical flares are labeled as ``0" to ``9" and sharp $\gamma$-rays flares as ``A" to ``J" (see text for details).      }
\label{plot_fig9a}
\end{figure*}

\subsection{Jet kinematics and broadband emission}
\label{vlbi_flare_corr} 
For the comparison of the observed broadband flaring behavior with the jet kinematics, we have used the flux density light 
curves from Paper I, where the details of observations and data reduction are given. Broadband flux light curves 
are shown in Fig.\ \ref{plot_fig9a}. Panel (a) presents the weekly averaged $\gamma$-ray photon flux light curve. 
Panel (b) shows the X-ray light curve, while panel (c) displays the optical V-band light curve. 
Panel (d) presents the radio (single-dish) and 43~GHz core\footnote{We used a mean value of the measured optically thin 
spectral index, $\alpha_{thin}$ = 0.4 \citep{rani2013a} to scale the 86~GHz flux density measurements to those at 43~GHz.} 
flux density curves plotted in the same panel. Panel (e) exhibit 
flux density light curves of the individual components shown in different colors, with the same color used to shade their respective
ejection times. The major outburst between JD' $\sim$840 and 1150 is marked with a horizontal arrow in panel (b). 
The repetitive optical flares with a typical duration of $\sim$120--140~days are 
label as ``0" to ``9", and the broadband flares are denoted ``G" for 
$\gamma$-rays, ``X" for X-rays, ``O" for optical and ``R" for radio followed by the number adjacent to them; e.g.,\ 
we have flares G0 to G9 at $\gamma$-rays, X6 at X-rays, O0 to O9 at optical, and R6 and R9 at radio frequencies. 
The sharp $\gamma$-ray flares  are labeled as ``A" to ``J", and have a duration of $\sim$10--20~days. 

Although VLBI 
observations of S5 0716+714 are missing at the peak of the two major radio flares (JD' $\sim$ 1150--1200 
and $\sim$1300), there is still
an apparent one-to-one correlation between the core flux and single-dish radio flux measurements. This 
suggests that the single-dish radio flares are produced in the core, which agrees with the 
high mean core-to-jet flux ratio, which ranges from 
1.3 to 21.4 with a mean of 4.9$\pm$4.0. The broadband flux cross-correlations (Paper I) also indicate 
that the flux
variations at $\gamma$-ray (and optical) and single-dish radio frequencies are correlated such 
that the peak of
the major optical/$\gamma$-ray outburst leads the peak of the radio outburst by a two-month time period.

\subsubsection{Component ejections and their connection with broadband flares}
Because of the superposition of different modes of flaring activity, the broadband flaring behavior is quite complex 
in the source. The major outburst (marked with a horizontal arrow in Fig. \ref{plot_fig9a}) is accompanied by faster 
repetitive optical/$\gamma$-ray flares (labeled as ``0" to ``9") and sharp $\gamma$-rays flares (labeled as ``A" to ``J"). 
The jet kinematic study during this period suggests significant changes in the jet morphology of the source; in total, 
seven new components were ejected. It is therefore plausible that the ejection of new components could be connected to the broadband flares. We have
identified a stationary feature at a distance of $\sim$0.15~mas from the core (Section 3.1.1). Interaction of a moving component 
with the stationary feature could amplify the magnetic field and accelerate particles to produce
a multi-waveband flare \citep{marscher2014, fromm2012}.

\begin{table}
\center
\caption{Component ejection epochs and broadband flares}
\begin{tabular}{l l l } \hline
Component  & \multicolumn{2}{c}{Flares observed at the time of} \\\hline
           & zero separation  &  component crosses  \\
           &from the core     & stationary feature \\\hline
NC1        &O0/G0     & G0B   \\
NC2        &O2/G2     & G4C   \\
NC3        &O3/G3     & O4D    \\
NC4        &O5E/G5E   & O6G/G6G  \\
NC5        &O5F/G5F   & G6I  \\
NC6        &O6/G6     & --  \\
NC7        &R6        & O9J/G9J \\\hline
\label{tab3}
\end{tabular}
\end{table}

In Fig.\ \ref{plot_fig9a}, the shaded areas represent the ejection 
times of new jet components and the green dashed lines mark the average epochs when a moving component crosses the 
stationary feature at $\sim$0.15~mas. In Table \ref{tab3}, we list the new components (col.~1), along with the flares 
that coincide with their respective ejection times (col.~2). Flares observed closer in time 
to the passage of a moving component across the stationary feature are listed in col.~3. 
We note that most of the sharp $\gamma$-ray flares (B, C, D, G, I, and J) are observed when a moving 
component crosses the stationary feature, while most of the repetitive faster flares (0, 2, 3, 5, and 6) coincide with 
the ejections of new components from the core.

In Paper I, we found that the high-energy (optical and $\gamma$-ray) emission is significantly correlated with 
the radio flares such that the major optical/$\gamma$-ray outburst propagates down to radio frequencies (R6 flare) with 
a time delay of $\sim$65~days, following a power-law dependence on frequency with a slope $\sim$0.3. We also noticed that the 
evolution of radio flare R6 is in accordance with a generalized shock model \citep{valtaoja1992}. This suggests 
that the origin of the major outburst could be related to a disturbance/shock that moves down the jet.

We interpret the broadband flaring events in the following manner (see Fig.\ \ref{plot_fig8}). 
A disturbance at the base of the jet could cause a shock wave to form and propagate down the 
jet \citep{valtaoja1992, marscher1985}, which could be related to the onset of the major optical/$\gamma$-ray outburst (JD' $\sim$840) 
and disturbs the jet outflow, resulting in the ejection of two new components (NC2 and NC3, Fig. \ref{plot_fig8}a). The core is 
optically thick at that point, as the shock forms upstream of the core; for this reason, we do not see much activity in 
the  core.  Repetitive optical/$\gamma$-ray flares and the curved trajectories of the associated components favor 
the propagation of the shock along a bent trajectory, possibly a helical path. The moving shock reaches its maximum 
strength at JD' $\sim$1150 (Fig. \ref{plot_fig8} b), close to 
the peak of the major optical/$\gamma$-ray outburst (see Section 3.3, Paper I for details), which also coincides with 
the onset of radio flare R6. 
At this point, the shock reaches the core, which starts to become optically thin. After this, further disturbances passing through the core correspond to the ejections of three new components (NC3, NC4, and NC5).  Major radio outburst R6 is observed $\sim$65~days 
later, accompanied by the ejection of component NC7 (Fig. \ref{plot_fig8} c).  
As the moving features cross the the stationary feature at $\sim$0.15~mas, at first, the stationary feature 
gets destructed after the interaction, and later it appears again. The times of passage of moving components  
through the stationary feature appear to coincide with the sharp $\gamma$-ray flares in most cases. 
Therefore, the sharp $\gamma$-ray flares seem to be produced via the interaction of moving features with the 
stationary one.

X-ray flare X6 coincides with a similar increase in the flux of the 43 GHz core. Optical flare O7 peaks about a month later, 
which roughly coincides with the expected passage through the stationary feature of the same disturbance that caused flare X6. 
However, no bright moving knot is apparent in the next VLBA image $\sim 20$ days after O7. Also, the dramatic decrease in 
the millimeter-wavelength flux during this interval implies that the disturbance is strongly quenched following flare O7. 
There are two anomalies in this sequence of events: there is no optical flare as the disturbance crosses the core, and there 
is no $\gamma$-ray activity during the entire time span covering flares X6 and O7. The first indicates that, for reasons that 
are unclear \citep[e.g., non-favorable magnetic field orientation relative to a shock;][]{summerlin2012}, electrons with energies 
high enough to produce optical and $\gamma$-ray emission were not produced in the core region at this time. The second implies 
that there was a dearth of seed photons for inverse-Compton scattering $\sim$0.15 mas from the core during flare O7.

  \begin{figure}
\includegraphics[scale=0.35,angle=0,trim=70 120 50 120, clip]{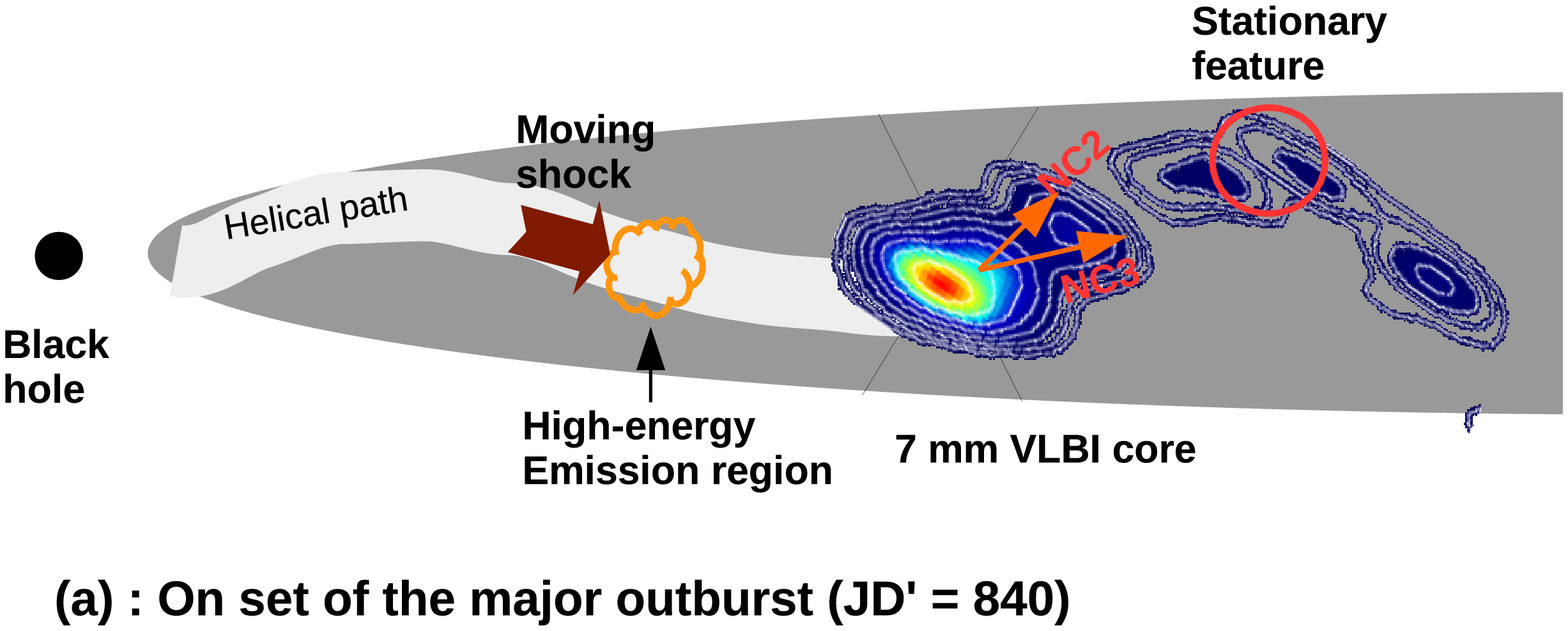}
\includegraphics[scale=0.35,angle=0,trim=70 120 50 120, clip]{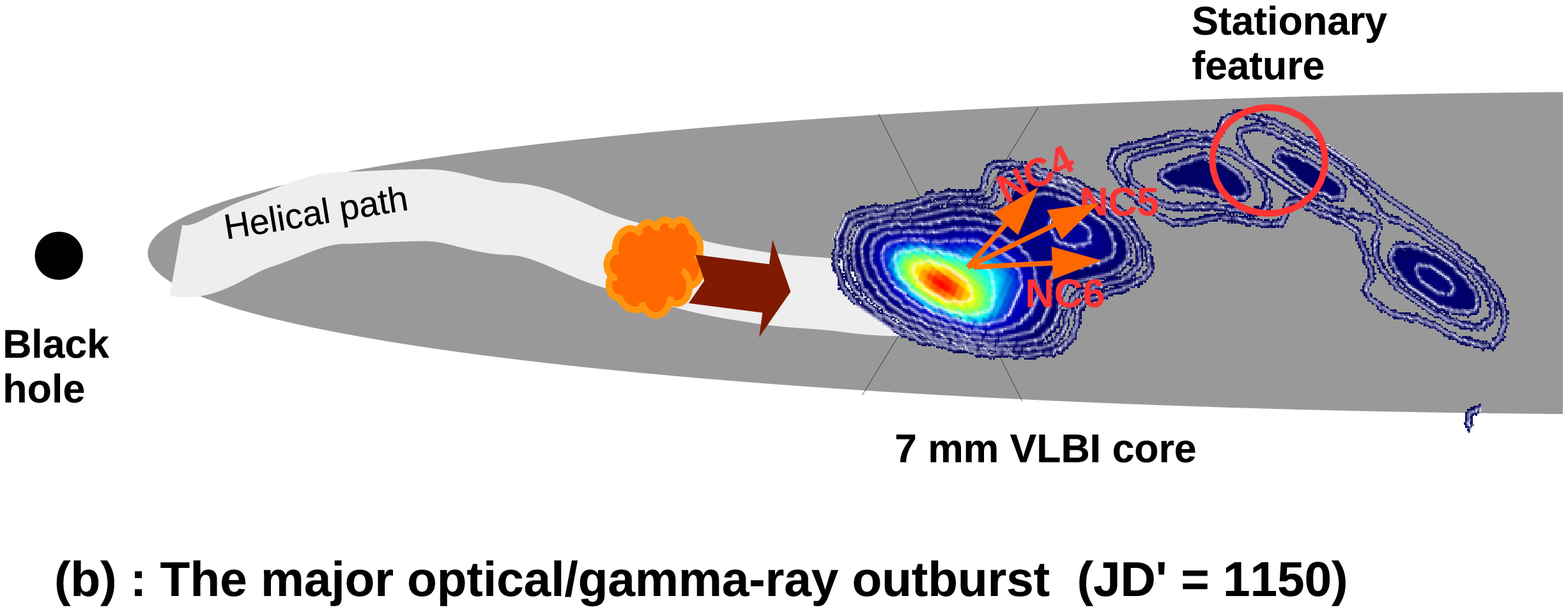}
\includegraphics[scale=0.35,angle=0,trim=70 120 50 120, clip]{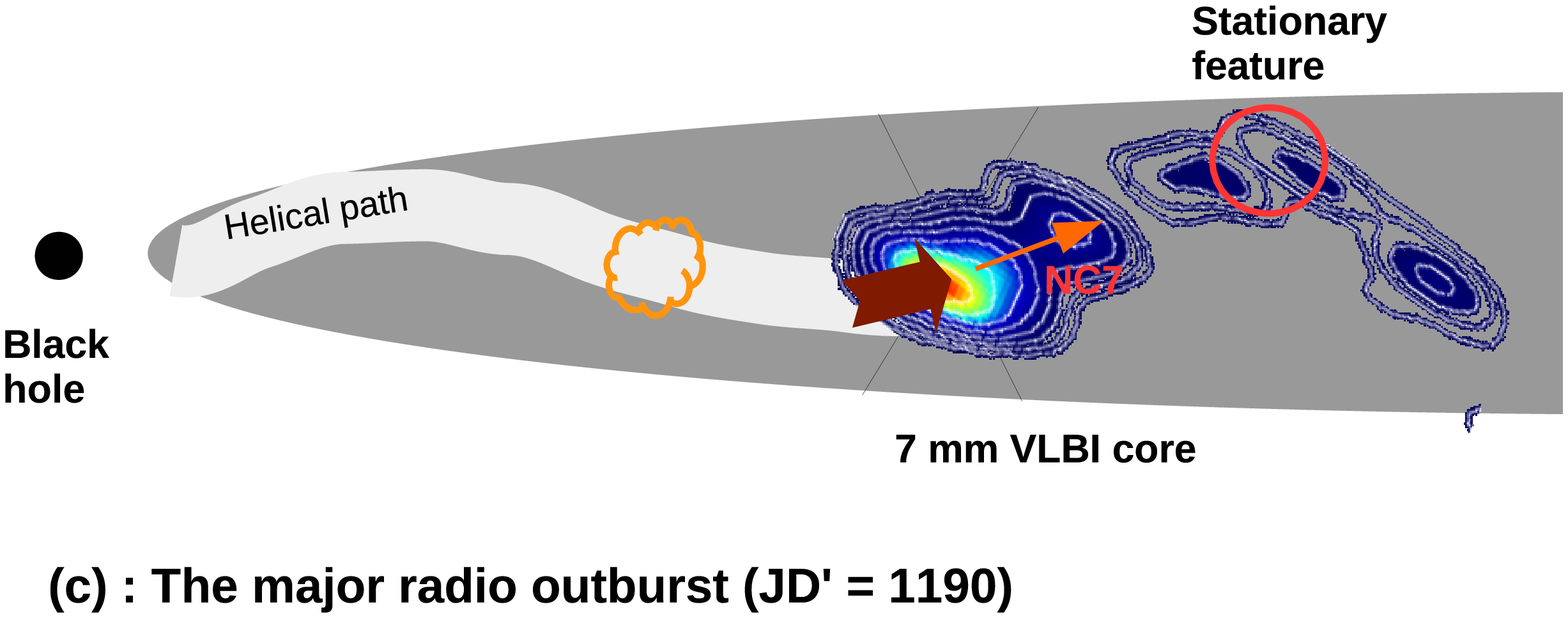}
  \caption{Sketch for the proposed scenario in the BL Lac S5 0716+714 (not to scale). 
Note that north is to the right in this sketch. A shock propagates down the jet along a helical 
path, producing a major optical/$\gamma$-ray outburst upstream of the 7~mm VLBI core and later a major 
radio outburst accompanied by ejection of components from the core. 
              }
\label{plot_fig8}
\end{figure}

\subsubsection{Doppler beaming of radio and gamma-ray emission regions}
The observed time lag between $\gamma$-ray and 43~GHz core flares can be used to estimate the separation between 
the two emission regions by using the following equation \citep{lars2014}: 
\begin{equation}
\Delta r_{\mathrm{\gamma,r}}=\frac{\beta_{\mathrm{app}}c\,\tau_{\mathrm{\gamma,r}}}{\sin\theta},
\end{equation} 
where $\theta$ is the viewing angle of the source, $\beta_{\mathrm{app}}$ is the apparent jet speed, and
$\tau_{\mathrm{\gamma,r}}$ is the observed time lag. Using the observed value $\tau_{\mathrm{\gamma,r}}$ = 82$\pm$32~days \citep{rani2014}, 
we have $\Delta r_{\mathrm{\gamma,r}}$ = (0.069$\pm$0.026)$\beta_{\mathrm{app}}$($\sin\theta$)$^{-1}$~pc. 

The measured apparent speeds 
in the source cover a wide range. The current observations suggest slower motion ($\leq$8~c) on sub-mas scales 
($\sim$0.3~mas), which later get accelerated to speeds as fast as 40~c on mas scales.  The source also has a prominent bent 
jet structure at sub-mas scales (see Fig. \ref{plot_fig2}); as a result, the sub-mas and mas regions could have different 
viewing angles. For $\beta_{\mathrm{app}}$ = 6-8~c, the viewing angle\footnote{$\theta = sin^{-1}(1/\beta_{\mathrm{app}})$}, 
$\theta$, is 6-9$^{\circ}$; this gives $\Delta r_{\mathrm{\gamma,r}}$ = 2.9-4.4~pc. Using  $\beta_{\mathrm{app}}$ = 40~c, we have 
$\theta$ = 1.4$^{\circ}$ and $\Delta r_{\mathrm{\gamma,r}}$ $\sim$ 110~pc. Even for a larger viewing angle, $\theta$=10$^{\circ}$, 
we obtain $\Delta r_{\mathrm{\gamma,r}}$ $\sim$16~pc. \citet{pushkarev2012} estimated 
that the 15~GHz core is located at a distance 
of 6.68~pc from the jet base/apex.  Using the technique to align the optically-thin emission based on model-fitted 
components and morphological similarities, we derive a shift of 0.042$\pm$0.013~mas (0.19$\pm$0.06~parsec) between the 15~GHz 
and 43~GHz core, which puts the location of the 43~GHz core at a distance of $\sim$6.5~pc from the jet-apex. 
Therefore, the distance between $\gamma$-ray emission region and 43~GHz core cannot be larger than 6.5~pc. 
For a typical viewing angle $\theta~\approx 5^{\circ}$, $\Delta r_{\mathrm{\gamma,r}}$ $\leq$ 6.5~pc gives  
$\beta_{\mathrm{app}}$ $\leq$ 8~c. However, we note that for a larger viewing angle ($\theta$ $\sim$ 10$^{\circ}$), 
$\beta_{\mathrm{app}}$ could be as high as 10-12$c$ but not $\sim 40c$ unless the viewing angle is larger than 30$^{\circ}$, which 
is unlikely to be the case for S5 0716+714. 
The results therefore suggest that the high-energy and radio emission 
regions have different apparent speeds and probably different viewing angles as well. This
implies that the $\gamma$-ray and 
radio emission regions have different Doppler factors.

\section{Summary and conclusion} 
The monthly sampled mm-VLBI monitoring of S5 0716+714 allows us to study the 
inner jet kinematics and to investigate its relation to the observed broadband flux variability. 
Our study reveals significant non-radial motions in the jet outflow, including variations in the orientation of the sub-parsec-scale
jet (see Rani et al. 2014 for details) and possibly helical trajectories of moving components.

In radial directions, the individual 
components exhibit extreme apparent speeds as high as $\sim$37~c.  An apparent speed of $\sim$20~c 
seems to be more typical in S5 0716+714 on parsec scales. Recently, \citet{lister2013} reported 
an apparent speed of 43.6$\pm$1.3~c. These values would even be extreme for a quasar. Therefore, moving components 
in the jets of BL Lac 
objects can be just as fast as in quasars, despite flux-limited surveys at lower resolution 
\citep[e.g.][]{lister2009} finding that the mean apparent velocity of BL Lac objects is lower than for 
quasars. Further mm-VLBI studies are needed to determine the extent to which angular resolution and 
the variety of spectral classes of BL Lac objects (low- to high-frequency peaks of the synchrotron spectral 
energy distributions) skew the apparent velocity distribution. It is possible that the fast BL Lacs 
are actually quasars in disguise, with their broad emission lines swamped by the non-thermal optical continuum. On the other hand, emission lines have never been observed in S5 0716+714.

The bright components in 0716+714 follow a continuous decay trend in the observed brightness 
temperature (T$_{B,obs} \propto r_{jet}^{-(2.36\pm0.41)}$), which implies a variation in Doppler factor 
along the jet axis. The spectral evolution of radio flares (see Paper I for details) also suggests 
variations in Doppler factor. A significant correlation between $\gamma$-ray flux and variations in orientation of the sub-parsec 
scale jet \citep{rani2014} supports the possibility that the observed flux variations in 
S5 0716+714 are strongly influenced by Doppler factor variations, and therefore may have a geometrical origin.    

Our study suggests a connection between jet kinematics and the observed broadband flaring activity. We 
find that the origin of major optical/$\gamma$-ray/radio outbursts in the source seems to be related to a disturbance, such as shock 
wave, propagating down the jet. The repetitive, faster optical/$\gamma$-ray flares, superposed on the 
major outbursts, and non-ballistic motions of the associated jet components can be interpreted via helical
jet models \citep[as discussed by, e.g.,][]{blandford1982, blandford1977, camenzind1992} and/or  
magnetohydrodynamic instabilities \citep{hardee2006}.  
Our analysis favors that the major optical/$\gamma$-ray flares in S5 0716+714 are 
produced upstream of the 7~mm VLBI core. The sharp $\gamma$-ray flares, however, seem to be 
produced when the moving components cross a stationary feature 0.15~mas from the core.
Finally, our analysis suggests that 
the upstream $\gamma$-ray emission region has a different Doppler factor than observed in
the parsec-scale radio jet.

\begin{acknowledgements}
The {\it Fermi}/LAT Collaboration acknowledges the generous support of a number of agencies
and institutes that have supported the {\it Fermi}/LAT Collaboration. These include the National
Aeronautics and Space Administration and the Department of Energy in the United States, the
Commissariat \`a l'Energie Atomique and the Centre National de la Recherche Scientifique / Institut
National de Physique Nucl\'eaire et de Physique des Particules in France, the Agenzia Spaziale
Italiana and the Istituto Nazionale di Fisica Nucleare in Italy, the Ministry of Education,
Culture, Sports, Science and Technology (MEXT), High Energy Accelerator Research Organization
(KEK) and Japan Aerospace Exploration Agency (JAXA) in Japan, and the K.\ A.\ Wallenberg
Foundation, the Swedish Research Council and the Swedish National Space Board in Sweden.

This research has made use of data obtained with the Global Millimeter VLBI Array (GMVA), which 
consists of telescopes operated by the MPIfR, IRAM, Onsala, Metsahovi, Yebes and the VLBA. The data 
were correlated at the correlator of the MPIfR in Bonn, Germany. The VLBA is an instrument of the 
National Radio Astronomy Observatory, a facility of the National Science Foundation operated under 
cooperative agreement by Associated Universities, Inc.
This study makes use of 43 GHz VLBA data from the
VLBA-BU Blazar Monitoring Program (VLBA-BU-BLAZAR; http://www.bu.edu/blazars/VLBAproject.html), funded
by NASA through the Fermi Guest Investigator Program. The VLBA is an instrument of the National Radio
Astronomy Observatory. The National Radio Astronomy Observatory is a facility of the National Science
Foundation operated by Associated Universities, Inc. The Submillimeter Array is a joint project between
the Smithsonian Astrophysical Observatory and the Academia Sinica Institute of Astronomy and Astrophysics 
and is funded by the Smithsonian Institution and the Academia Sinica. BR is thankful to Svetlana Jorstad, 
Benoit Lott, Biagina Boccardi, and Shoko Koyama for useful discussion and comments on the paper. We 
thank the referee for constructive comments.
\end{acknowledgements}


\begin{appendix}
\section{Model fit component results }
\begin{table*}[h]
\caption{Results from Gaussian Model fitting and component parameters }
\begin{tabular}{c c c c c c c }\hline
Epoch  (JD$^{\prime}$)       & S$_{peak}$        & r               & $\theta$           & $\phi$  & Comp$^{a}$ & Wavelength  \\ 
 Date & (Jy/beam)         &(mas)            & ($^{\circ}$)       & (mas)   &    &  \\\hline
720  & 1.670$\pm$0.084  &  0$\pm$0   &    0$\pm$0      & 0.044$\pm$0.002  &  core & 7~mm \\       
10.09.2008     & 0.139$\pm$0.007  &  0.098$\pm$0.005   &   39.90$\pm$2.00      & 0.034$\pm$0.002  &  A1 &   \\       
     & 0.023$\pm$0.001  &  0.714$\pm$0.036   &   13.28$\pm$0.66   & 0.427$\pm$0.021  &  C1  &  \\       
787  & 1.670$\pm$0.084  &  0$\pm$0   &    0$\pm$0      & 0.038$\pm$0.002  &  core &7~mm \\      
16.11.2008     & 0.188$\pm$0.009  &  0.103$\pm$0.005   &   29.29$\pm$1.46   & 0.071$\pm$0.004  &  A1 &    \\       
     & 0.022$\pm$0.001  &  1.516$\pm$0.076   &   12.33$\pm$0.62   & 0.487$\pm$0.024  &  C1  &   \\     
822  & 2.044$\pm$0.102  &  0$\pm$0   &    0$\pm$0      & 0.034$\pm$0.002  &  core 7~mm  \\
21.12.2008     & 0.082$\pm$0.004  &  0.139$\pm$0.007   &   26.83$\pm$1.34   & 0.101$\pm$0.005  &  NC1&    \\
     & 0.013$\pm$0.001  &  1.700$\pm$0.085   &   13.57$\pm$0.68   & 0.538$\pm$0.027  &  C1   &  \\
856  & 2.167$\pm$0.108  &  0$\pm$0   &    0$\pm$0      & 0.034$\pm$0.002  &  core  7~mm \\
24.01.2009     & 0.405$\pm$0.020  &  0.108$\pm$0.005   &    8.06$\pm$0.40   & 0.045$\pm$0.002  &  A1 &    \\
     & 0.052$\pm$0.003  &  0.292$\pm$0.015   &   15.92$\pm$0.80   & 0.171$\pm$0.009  &  NC1   & \\
     & 0.025$\pm$0.001  &  1.572$\pm$0.079   &   11.01$\pm$0.55   & 0.879$\pm$0.044  &  C1    & \\
885  & 1.717$\pm$0.086  &  0$\pm$0   &       0$\pm$0      & 0.035$\pm$0.002  &  core  & 7~mm \\
22.02.2009     & 0.420$\pm$0.021  &  0.125$\pm$0.006   &    6.64$\pm$0.33   & 0.051$\pm$0.003  &  A1 &     \\
     & 0.053$\pm$0.003  &  0.383$\pm$0.019   &    9.89$\pm$0.49   & 0.201$\pm$0.010  &  NC1 &   \\
     & 0.016$\pm$0.001  &  1.873$\pm$0.094   &   13.47$\pm$0.67   & 0.642$\pm$0.032  &  C1  &   \\
924  & 1.038$\pm$0.052  &  0$\pm$0   &       0$\pm$0      & 0.018$\pm$0.001  &  core &7~mm   \\
02.04.2009     & 0.552$\pm$0.028  &  0.036$\pm$0.002   &   56.28$\pm$2.81   & 0.036$\pm$0.002  &  NC2 &    \\
     & 0.187$\pm$0.009  &  0.150$\pm$0.008   &   13.26$\pm$0.66   & 0.062$\pm$0.003  &  A1    & \\
     & 0.051$\pm$0.003  &  0.460$\pm$0.023   &    7.60$\pm$0.38   & 0.163$\pm$0.008  &  NC1   & \\
     & 0.020$\pm$0.001  &  1.672$\pm$0.084   &    8.84$\pm$0.44   & 0.692$\pm$0.035  &  C1    & \\
963  & 0.354$\pm$0.018  &  0$\pm$0   &       0$\pm$0      & 0.011$\pm$0.001  &  core  &3~mm \\
11.05.2009     & 0.104$\pm$0.005  &  0.028$\pm$0.001   &   73.33$\pm$3.67   & 0.012$\pm$0.001  &  NC3 &    \\
     & 0.043$\pm$0.002  &  0.084$\pm$0.004   &   47.52$\pm$2.38   & 0.025$\pm$0.001  &  NC2   & \\
     & 0.048$\pm$0.002  &  0.159$\pm$0.008   &   54.57$\pm$2.73   & 0.030$\pm$0.001  &  A1    & \\
     & 0.030$\pm$0.001  &  0.441$\pm$0.022   &   16.93$\pm$0.85   & 0.047$\pm$0.002  &  K1    & \\
     & 0.019$\pm$0.001  &  0.628$\pm$0.031   &   39.92$\pm$2.00   & 0.044$\pm$0.002  &  NC1   & \\
982  & 0.482$\pm$0.024  &  0$\pm$0   &       0$\pm$0      & 0.038$\pm$0.002  &  core  &7~mm \\
30.05.2009     & 0.118$\pm$0.006  &  0.109$\pm$0.005   &   29.31$\pm$1.47   & 0.074$\pm$0.004  &  NC2&    \\
     & 0.035$\pm$0.002  &  0.525$\pm$0.026   &    5.61$\pm$0.28   & 0.192$\pm$0.010  &  K1    & \\
1004 & 0.548$\pm$0.027  &  0$\pm$0   &       0$\pm$0      & 0.039$\pm$0.002  &  core  &7~mm \\
21.06.2009     & 0.075$\pm$0.004  &  0.104$\pm$0.005   &   31.10$\pm$1.56   & 0.097$\pm$0.005  &  NC3 &    \\
     & 0.022$\pm$0.001  &  0.594$\pm$0.030   &    8.26$\pm$0.41   & 0.148$\pm$0.007  &  K1     & \\
     & 0.564$\pm$0.028  &  0$\pm$0   &       0$\pm$0      & 0.035$\pm$0.002  &  core  &7~mm \\
1040 & 0.166$\pm$0.008  &  0.080$\pm$0.004   &   39.50$\pm$1.98   & 0.040$\pm$0.002  &  A2    & \\
27.07.2009     & 0.038$\pm$0.002  &  0.165$\pm$0.008   &   19.10$\pm$0.96   & 0.067$\pm$0.003  &  NC3 &    \\
     & 0.010$\pm$0.001  &  0.690$\pm$0.035   &   11.10$\pm$0.56   & 0.187$\pm$0.009  &  K1    & \\
1060 & 0.663$\pm$0.033  &  0$\pm$0   &       0$\pm$0      & 0.039$\pm$0.002  &  core  &7~mm \\
16.08.2009     & 0.125$\pm$0.006  &  0.079$\pm$0.004   &   56.40$\pm$2.82   & 0.040$\pm$0.002  &  A2 &     \\
     & 0.030$\pm$0.002  &  0.188$\pm$0.009   &   23.50$\pm$1.18   & 0.066$\pm$0.003  &  NC3   & \\
     & 0.019$\pm$0.001  &  0.494$\pm$0.025   &    6.77$\pm$0.34   & 0.181$\pm$0.009  &  NC2   & \\
     & 0.006$\pm$0.000  &  1.098$\pm$0.055   &    2.52$\pm$0.13   & 0.104$\pm$0.005  &  NC1   & \\
1091 & 0.873$\pm$0.044  &  0$\pm$0   &       0$\pm$0      & 0.045$\pm$0.002  &  core  &7~mm \\
16.09.2009     & 0.172$\pm$0.009  &  0.112$\pm$0.006   &   48.70$\pm$2.44   & 0.044$\pm$0.002  &  NC4 &    \\
     & 0.041$\pm$0.002  &  0.388$\pm$0.019   &   11.50$\pm$0.58   & 0.199$\pm$0.010  &  NC3   & \\
     & 0.015$\pm$0.001  &  0.764$\pm$0.038   &    9.97$\pm$0.50   & 0.147$\pm$0.007  &  NC2   & \\
1117 & 0.356$\pm$0.018  &  0$\pm$0   &       0$\pm$0      & 0.012$\pm$0.001  &  core  &3~mm \\
12.10.2009     & 0.110$\pm$0.005  &  0.029$\pm$0.001   &   73.71$\pm$3.69   & 0.015$\pm$0.001  &  NC6 &    \\
     & 0.041$\pm$0.002  &  0.084$\pm$0.004   &   48.90$\pm$2.45   & 0.017$\pm$0.001  &  NC5   & \\
     & 0.042$\pm$0.002  &  0.158$\pm$0.008   &   53.49$\pm$2.67   & 0.024$\pm$0.001  &  NC4   & \\
     & 0.021$\pm$0.001  &  0.502$\pm$0.025   &   17.16$\pm$0.86   & 0.040$\pm$0.002  &  NC3   & \\
1119 & 0.835$\pm$0.042  &  0$\pm$0   &       0$\pm$0      & 0.039$\pm$0.002  &  core  &7~mm \\
14.10.2009     & 0.122$\pm$0.006  &  0.100$\pm$0.005   &   47.00$\pm$2.35   & 0.041$\pm$0.002  &  NC5&    \\
     & 0.025$\pm$0.001  &  0.563$\pm$0.028   &   11.50$\pm$0.58   & 0.197$\pm$0.010  &  NC3   & \\
     & 0.013$\pm$0.001  &  0.875$\pm$0.044   &   15.00$\pm$0.75   & 0.189$\pm$0.009  &  NC2   & \\
1121 & 1.142$\pm$0.057  &  0$\pm$0   &       0$\pm$0      & 0.040$\pm$0.002  &  core  &7~mm \\
16.10.2009     & 0.175$\pm$0.009  &  0.100$\pm$0.005   &   48.10$\pm$2.41   & 0.041$\pm$0.002  &  NC5&    \\
     & 0.038$\pm$0.002  &  0.607$\pm$0.030   &   10.30$\pm$0.52   & 0.248$\pm$0.012  &  NC3   & \\
     & 0.007$\pm$0.000  &  1.201$\pm$0.060   &   20.30$\pm$1.02   & 0.352$\pm$0.018  &  NC1   & \\
1125 & 1.204$\pm$0.060  &  0$\pm$0   &       0$\pm$0      & 0.039$\pm$0.002  &  core  &7~mm \\
20.10.2009     & 0.243$\pm$0.012  &  0.103$\pm$0.005   &   43.20$\pm$2.16   & 0.052$\pm$0.003  &  NC5 &    \\
     & 0.036$\pm$0.002  &  0.648$\pm$0.032   &    9.79$\pm$0.49   & 0.238$\pm$0.012  &  NC3   & \\
     & 0.010$\pm$0.001  &  1.177$\pm$0.059   &    9.19$\pm$0.46   & 0.391$\pm$0.020  &  NC1   & \\\hline
\end{tabular} \\
JD$^{\prime}$ = JD - 2454000, \\
S$_{peak}$ : The integrated flux in the component, \\
r : The radial distance of the component center from the center of the map, \\
$\theta$ : The position angle of the center of the component, and \\ 
$\phi$ : The FWHM of the component. \\
\end{table*}

\begin{table*}
\begin{tabular}{c c c c c c c }
{Table 1 continued. } \\\hline
Epoch (JD$^{\prime}$)        & S$_{peak}$        & r               & $\theta$           & $\phi$  & Comp$^{a}$ & Wavelength  \\ 
Date  & (Jy/beam)         &(mas)            & ($^{\circ}$)       & (mas)   &   &  \\\hline
1131 & 1.446$\pm$0.072  &  0$\pm$0   &       0$\pm$0      & 0.035$\pm$0.002  &  core &7~mm   \\
25.10.2009     & 0.262$\pm$0.013  &  0.115$\pm$0.006   &   40.70$\pm$2.04   & 0.052$\pm$0.003  &  NC5 &    \\
     & 0.029$\pm$0.001  &  0.675$\pm$0.034   &   10.90$\pm$0.55   & 0.224$\pm$0.011  &  NC3  &  \\
     & 0.013$\pm$0.001  &  1.256$\pm$0.063   &   14.30$\pm$0.72   & 0.227$\pm$0.011  &  NC1  &  \\
1152 & 4.144$\pm$0.207  &  0$\pm$0   &       0$\pm$0      & 0.032$\pm$0.002  &  core &7~mm  \\
16.11.2009     & 0.397$\pm$0.020  &  0.075$\pm$0.004   &   31.60$\pm$1.58   & 0.064$\pm$0.003  &  NC6 &    \\
     & 0.077$\pm$0.004  &  0.254$\pm$0.013   &   27.30$\pm$1.37   & 0.099$\pm$0.005  &  NC4    & \\
     & 0.018$\pm$0.001  &  0.736$\pm$0.037   &   10.70$\pm$0.54   & 0.214$\pm$0.011  &  NC3   & \\
1207 & 1.880$\pm$0.094  &  0$\pm$0   &       0$\pm$0      & 0.043$\pm$0.002  &  core  &7~mm \\
10.01.2010     & 0.206$\pm$0.010  &  0.108$\pm$0.005   &   20.30$\pm$1.02   & 0.060$\pm$0.003  &  NC6 &    \\
     & 0.025$\pm$0.001  &  0.380$\pm$0.019   &   17.20$\pm$0.86   & 0.224$\pm$0.011  &  NC4   & \\
     & 0.015$\pm$0.001  &  0.817$\pm$0.041   &    9.89$\pm$0.49   & 0.271$\pm$0.014  &  NC3   & \\
1239 & 1.503$\pm$0.075  &  0$\pm$0   &       0$\pm$0      & 0.035$\pm$0.002  &  core  &7~mm \\
11.02.2010     & 0.219$\pm$0.011  &  0.046$\pm$0.002   &   48.90$\pm$2.45   & 0.041$\pm$0.002  &  NC7 &    \\
     & 0.064$\pm$0.003  &  0.138$\pm$0.007   &   15.50$\pm$0.78   & 0.072$\pm$0.004  &  NC6   & \\
     & 0.031$\pm$0.002  &  0.355$\pm$0.018   &   11.90$\pm$0.60   & 0.280$\pm$0.014  &  NC5   & \\
     & 0.007$\pm$0.000  &  0.951$\pm$0.048   &   12.07$\pm$0.60   & 0.152$\pm$0.008  &  NC3   & \\
1262 & 1.453$\pm$0.073  &  0$\pm$0   &       0$\pm$0      & 0.035$\pm$0.002  &  core  &7~mm \\
06.032010     & 0.477$\pm$0.024  &  0.082$\pm$0.004   &   38.90$\pm$1.94   & 0.041$\pm$0.002  &  NC7 &    \\
     & 0.050$\pm$0.002  &  0.198$\pm$0.010   &    6.09$\pm$0.30   & 0.072$\pm$0.004  &  NC6   & \\
     & 0.016$\pm$0.001  &  0.550$\pm$0.027   &   12.60$\pm$0.63   & 0.254$\pm$0.013  &  NC4   & \\
     & 0.011$\pm$0.001  &  1.464$\pm$0.073   &    7.88$\pm$0.39   & 0.374$\pm$0.019  &  NC3   & \\
1327 & 0.902$\pm$0.045  &  0$\pm$0   &       0$\pm$0      & 0.020$\pm$0.001  &  core  &3~mm \\
10.05.2010     & 0.194$\pm$0.010  &  0.065$\pm$0.003   &   71.62$\pm$3.58   & 0.032$\pm$0.002  &  A3 &     \\
     & 0.019$\pm$0.001  &  0.560$\pm$0.028   &   14.54$\pm$0.73   & 0.024$\pm$0.001  &  NC7   & \\
     & 0.213$\pm$0.011  &  0.134$\pm$0.007   &   70.29$\pm$3.51   & 0.063$\pm$0.003  &  X     & \\
     & 0.023$\pm$0.001  &  0.200$\pm$0.010   &   45.00$\pm$2.25   & 0.021$\pm$0.001  &  NC5   & \\
1336 & 1.543$\pm$0.077  &  0$\pm$0   &       0$\pm$0      & 0.041$\pm$0.002  &  core  &7~mm \\
19.05.2010     & 0.359$\pm$0.018  &  0.109$\pm$0.005   &   58.30$\pm$2.92   & 0.062$\pm$0.003  &  A3 &     \\
     & 0.043$\pm$0.002  &  0.435$\pm$0.022   &   12.40$\pm$0.62   & 0.232$\pm$0.012  &  NC6   & \\
     & 0.009$\pm$0.000  &  1.624$\pm$0.081   &   15.20$\pm$0.76   & 0.323$\pm$0.016  &  X     & \\
1362 & 0.783$\pm$0.039  &  0.000$\pm$0.000   &       0$\pm$0      & 0.041$\pm$0.002  &  core  &7~mm \\
14.06.2010     & 0.261$\pm$0.013  &  0.110$\pm$0.005   &   49.50$\pm$2.48   & 0.075$\pm$0.004  &  A3 &     \\
     & 0.020$\pm$0.001  &  0.687$\pm$0.034   &   12.00$\pm$0.60   & 0.251$\pm$0.013  &  NC6   & \\
     & 0.005$\pm$0.000  &  1.781$\pm$0.089   &   15.60$\pm$0.78   & 0.171$\pm$0.009  &  X     & \\
1410 & 1.389$\pm$0.069  &  0.000$\pm$0.000   &       0$\pm$0      & 0.040$\pm$0.002  &  core  &7~mm \\
01.08.2010     & 0.438$\pm$0.022  &  0.070$\pm$0.003   &   70.20$\pm$3.51   & 0.053$\pm$0.003  &  A3 &     \\
     & 0.080$\pm$0.004  &  0.250$\pm$0.013   &   29.70$\pm$1.49   & 0.173$\pm$0.009  &  NC7   & \\
     & 0.014$\pm$0.001  &  0.620$\pm$0.031   &    8.66$\pm$0.43   & 0.181$\pm$0.009  &  NC6   & \\
1458 & 2.974$\pm$0.149  &  0$\pm$0   &       0$\pm$0      & 0.033$\pm$0.002  &  core  &7~mm \\
18.09.2010     & 0.535$\pm$0.027  &  0.081$\pm$0.004   &   73.60$\pm$3.68   & 0.040$\pm$0.002  &  A3 &     \\
     & 0.030$\pm$0.002  &  0.185$\pm$0.009   &   50.50$\pm$2.53   & 0.050$\pm$0.003  &  X      & \\
     & 0.022$\pm$0.001  &  0.324$\pm$0.016   &   29.80$\pm$1.49   & 0.176$\pm$0.009  &  NC7    & \\
     & 0.032$\pm$0.002  &  0.693$\pm$0.035   &   10.90$\pm$0.55   & 0.235$\pm$0.012  &  NC6    & \\\hline
\end{tabular} \\
$a$ :  Identification of the individual components. If a component appeared only in a single epoch, 
we labeled it with X.  \\
\label{tab1}
\end{table*}

\end{appendix}

\end{document}